# Quantum Mirrors and Crossing Symmetry as Heart of Ghost Imaging

D. B. Ion[1)], M. L. Ion[2)] and L. Rusu[1)]

[1)]*Institute for Physics and Nuclear Engineering, Department of Fundamental Physics, POB MG 6, Bucharest, Romania* [2)]*Bucharest University, Department for Nuclear Physics, POB MG 11, Bucharest, Romania*

**Abstract**: In this paper it is proved that the key to understanding the ghost imaging mystery are the crossing symmetric photon reactions in the nonlinear media. Hence, the laws of the plane quantum mirror (QM) and that of spherical quantum mirror, observed in the ghost imaging experiments, are obtained as natural consequences of the energy-momentum conservation laws. So, it is shown that the ghost imaging laws depend only on the energy-momentum conservation and not on the photons entanglement. The extension of these results to the ghost imaging with other kind of light is discussed. Some fundamental experiments for a decisive tests of the [SPDC-DFG]-quantum mirror are suggested.

## 1. Introduction

It is well known that the *crossing symmetry of S-matrix* is a fundamental theorem of quantum field theory [1,2]. So, the crossing symmetry is one of the most important practical elements of dynamics which makes use of the analytical properties of the scattering amplitudes [3]. It relates various amplitudes, for example helicity amplitudes, in one channel to those in other channels, or more generally to other processes in which one or more or all incoming and outgoing particles have been interchanged. Roughly speaking, the processes in crossed channels provide forces for the process in the original channel. For example, forces responsible for the binding of two particles may be attributed to the exchange of the other particles in crossed channels. As is well known, the crossing relations generalize Pauli exchange principle.

Quantum electrodynamics (QED) being one of the working example of the field theory is indeed crossing symmetric. This property of the QED is experimentally verified [3] with very high accuracy. The S-matrix crossing symmetry [2] is a symmetry that relates S-matrix elements. Interaction processes involving different kinds of particles can be obtained from each other by replacing incoming particles with outgoing antiparticles after taking the analytic continuation. The crossing symmetry applies to all known particles, including the photon which is its own antiparticle. For example, the annihilation of an electron with a positron into two photons is related to an elastic scattering of an electron with a photon by crossing symmetry. This relation allows to calculate the scattering amplitude of one process from the amplitude for the other process if negative values of energy of some particles are substituted

$$\gamma + e^- \rightarrow e^- + \gamma \qquad (1)$$

$$e^- + e^+ \rightarrow \gamma + \gamma \qquad (2)$$

By examination, it can be seen that these two interactions are related by crossing symmetry. It could then be said that the observation of Compton scattering implies the existence of pair annihilation and predicts that it will produce a pair of photons. Here some remarks are necessary in connection with the quantum entanglement. If the quantum entanglement is a quantum mechanical phenomenon in which the quantum states of two or more objects have to be described with reference to each other, even though the individual objects may be spatially separated, then the crossing symmetry of an interaction can be interpreted as a very special kind of entanglement (see M. Pitkanen in Ref. [4]).

On the other hand, in 1995, at Baltimore University, professor (Dr.) Yanhua Shih initiated ghost-imaging research [5], by using entangled photons. In that experiment, one photon passed through stenciled patterns in a mask to trigger a detector, and another photon was captured by a second detector. Surprisingly, an image of the pattern between the two detectors appeared which the physics community called ghost-imaging. Some definitions of ghost imaging (see refs. [6-8]) are as follows:
(i) Ghost-imaging is a visual image of an object created by means of light that has never interacted with the object. (ii) Ghost imaging is an unusual effect by which an image is formed using light patterns that do not emanate from the target object. (iii) Ghost imaging, is a novel technique in which the object and the image system are on separate optical paths. (iv) "Ghost-imaging is similar to taking a flash-lit photo of an object using a normal camera. The image is formed by the photons that come out of the flash, bounce off the object, and then are focused through the lens onto photo-reactive film or a charge-coupled array. But, in this case, the image is not formed from light that hits the object and bounces back," Dr. Shih said. "The camera collects photons from the light sources that did not hit the object, but are paired through a quantum effect with others that did".

Here, we must underline that ten years ago, based on the crossing symmetry of the SPDC-photon reactions, the authors of the papers [9-11] presented a new explanation of all the ghost phenomena. Then, they introduced the concept of SPDC-quantum mirror (QM) and on this basis they proved some important Quantum Mirror physical laws (see also our paper [12]) which can be of great help for a more deep understanding of the ghost imaging phenomena.

This article is a direct continuation of our paper [12]. So, in Sect. 2 we present the crossing symmetric photon reactions as well as the connection between crossing symmetry and difference frequency generation (DFG). Moreover, in Sect. 2, some theoretical results for the DFG-conversion rate, in three wave mixing approximation, are presented. In Sect. 3 ghost imaging via [SPDC, DFG]-quantum mirrors are described and experimentally verified. Sect. 4 is reserved for discussions and conclusions.

## 2. Crossing symmetric [SPDC-DFG]-photon reactions

The first experiments on ghost imaging were performed using a pair of entangled photons produced by spontaneous parametric down conversion (SPDC). In this process, a primary pump (p) photon is incident on a nonlinear crystal and produces the photons idler (i) and signal (s) by the reaction: $p \to s + i$. These photons are correlated in energy, momentum, polarizations and time of birth. Some of these features, such as energy and momentum conservations: $\omega_p = \omega_s + \omega_i$, $k_p = k_s + k_i$ are exploited to match in diverse experiments. (e.g., Momentum conservation in the "degenerate" case when the idler and the signal photons acquire the same frequency leads to the production of a pair of simultaneous photons that are emitted at equal angles relative to the incident beam). Now, if the S-matrix crossing symmetry [1-3] of the electromagnetic interaction in the spontaneous parametric down conversion (SPDC) crystals is taken into account, then the existence of the direct SPDC process (see Fig.1)

$$\gamma_p \to \gamma_s + \gamma_i \qquad (3)$$

will imply the existence of the following crossing symmetric processes

$$\gamma_p + \overline{\gamma_s} \to \gamma_i \qquad (4)$$

$$\gamma_p + \overline{\gamma_i} \to \gamma_s \qquad (5)$$

as real processes which can be described by the same transition amplitude. Here, by $\overline{s}$ and $\overline{i}$ we denoted the time reversed photons relative to the original photons s and i, respectively.

In fact, the SPDC-reactions (4)-(5) can be identified as being directly connected with the $\chi^{(2)}$ — second-order nonlinear effects called in general three waves mixing. So, the process (3) can be considered as being just the inverse of second-harmonic generation, while, the effects (4)-(5) corresponds to the *difference-frequency generation* (DFG) in the presence of pump laser via three wave mixing (see Fig.1). Here we have the generation of a photon at frequency $\omega_i$ when photons at frequencies $\omega_p$ and $\omega_s$ are incident on a crystal with an appreciable second-order susceptibility $\chi^{(2)}$, such that $\omega_i = \omega_p - \omega_s$ (assuming $\omega_p > \omega_s$). In order for energy conservation to hold, this additionally implies that, for every photon generated at the difference frequency $\omega_i$, a photon at $\omega_s$ must also be created, while a photon at the higher frequency $\omega_p$ must be annihilated. In addition, for the process to occur with an appreciable efficiency of frequency conversion, phase-matching must occur, so that: $\vec{k}_i = \vec{k}_p - \vec{k}_s$.

In many situations, the field at $\omega_p$ is an *intense pump field*, while the field at $\omega_s$ is a weak *signal field*. Difference-frequency generation yields amplification of the $\omega_s$ field (along with generation of another field at $\omega_i$, commonly called the *idler field*). Thus, this process is termed parametric amplification; when $\omega_s = \omega_i$, the device created is called *Degenerate Parametric Amplifier*. In fact, the SPDC-reactions (4)-(5) can be identified as being directly connected with the $\chi^{(2)}$ −second-order nonlinear effects called in general three waves mixing. So, the process (3) can be considered as being just the inverse of second-harmonic generation, while, the effects (4)-(5) corresponds to the *difference-frequency generation* (DFG) in the presence of pump laser via three wave mixing (see Fig. 1 ).

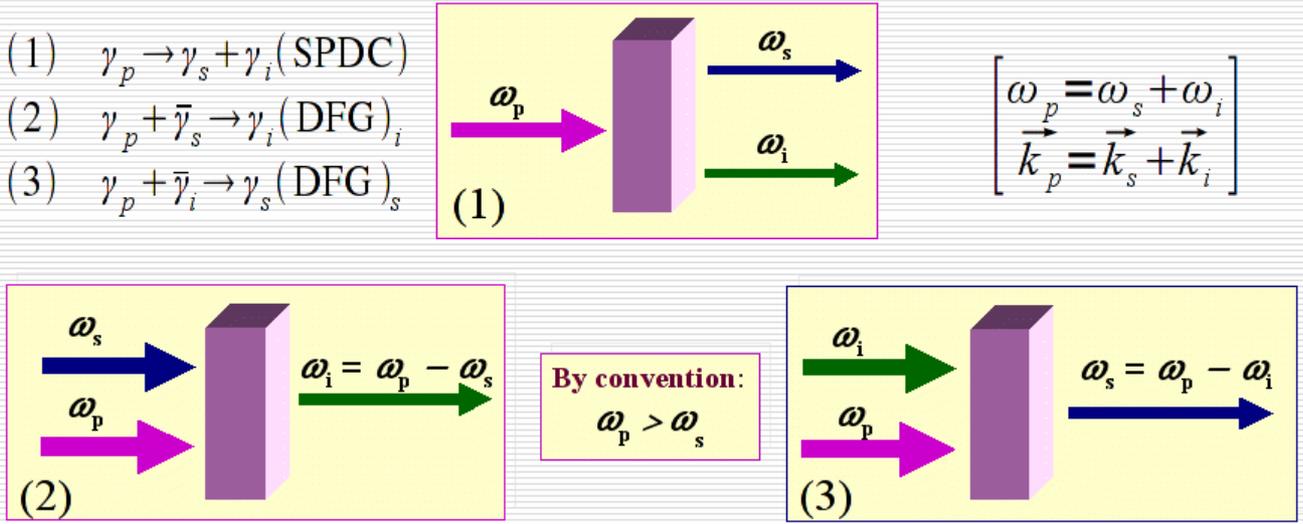

Fig.1: Schematic description of the crossing symmetry in electromagnetic transitions

Difference-frequency generation as a three-wave nonlinear interaction process has been theoretically analyzed by Armstrong et al. [13] in 1962. The difference-frequency conversion efficiency has been investigated by Boyd and Kleinman [14], based on the electric field generated by two focused Gaussian beams, studying the dependence of the generation power on the focusing condition and the properties of the nonlinear mixing material.

Therefore, let us consider two collinear Gaussian beams [called pump (p) and signal (s), with the powers $P_p$ and $P_s$ at frequencies $\omega_p$ and $\omega_s$, for the pump and signal beam, respectively] having identical confocal parameters $b = k_p w_p^2 = k_s w_s^2$ (here $w$ is the beam waist). In this case, the DFG output power $P_i$, at the difference-frequency $\omega_i = \omega_p - \omega_s$, can be written as follows [15,16]:

$$P_i = \frac{256\pi^2}{c^3} \cdot \omega_i \cdot \frac{(d_{\text{eff}})^2}{n_i \cdot n_s \cdot n_p} \cdot \frac{h(\mu, \xi, \alpha)}{(k_s^{-1} + k_p^{-1})} \cdot L \cdot P_p \cdot P_s \cdot \exp(-\alpha L) \qquad (6)$$

where $c$ is the speed of light in vacuum, $n$ is the index of refraction, $L$ is the crystal length, $d_{eff}$ is the effective nonlinear coefficient, and $\alpha$ is the absorption coefficient of the nonlinear optical medium at the DFG frequency. The subscripts $s, p, i$ refer to the signal, pump, and idler (infrared) photons, respectively. The focusing function $h(\mu, \xi, \alpha)$ involving walk-off and focused beam effects is given as (focusing point is assumed at the center of the crystal):

$$h(\mu, \xi, \alpha) = \frac{1}{4\cdot\xi} \cdot \int_{-\xi}^{+\xi} d\tau \int_{-\xi}^{+\xi} d\tau' \frac{\exp\left[\frac{b\cdot\alpha}{\xi}(\tau-\tau')\right]}{1 - \frac{j}{2}\cdot\left[\frac{1+\mu}{1-\mu} + \frac{1-\mu}{1+\mu}\right]\cdot(\tau-\tau') + \tau\cdot\tau'} \qquad (7)$$

where $\mu = k_p/k_s$ and the focusing parameter $\xi = L/b$ which relates the crystal length to the beam size of the pump and signal.

Difference frequency generation provides mid-infrared laser radiation by means of interaction of two near-infrared lasers in a non-linear medium. In connection with DFG-lasers, recently Stry et al. [18] developed a continuously mode-hop free tunable difference DFG-laser system suitable for high-resolution spectroscopy in the 3 micron region (see Fig.2 for a short description)

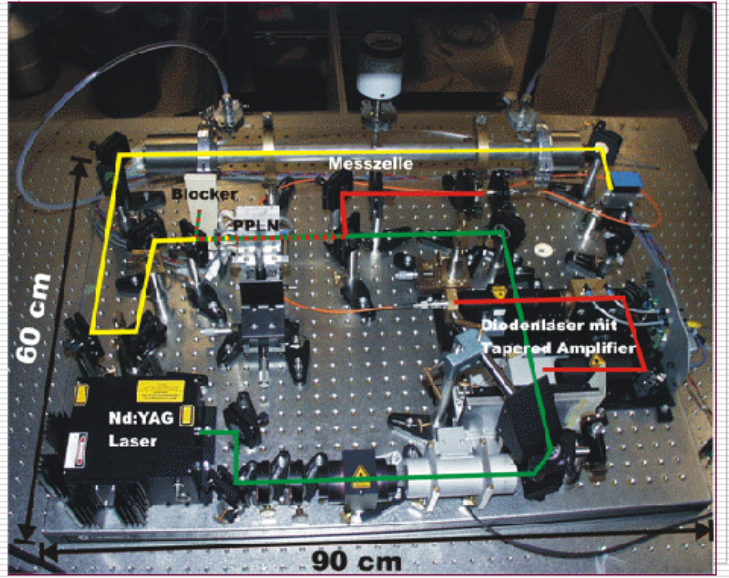

Fig.2: A short description of the portable DFG-laser of S. Stry, P. Hering, M. Mürtz [18]

## 3. Ghost imaging via [SPDC-DFG] crossing symmetric photon reactions

The main purpose here is to obtain an answer to the basic question: Is ghost imaging mystery solved via the quantum mirror (QM) mechanism introduced in refs. [9-12]?

So, we start with the definition of the quantum mirror and some of its physical laws [9-12].

*[SPDC,DFG]-Quantum Mirror (QM)* (see Fig.3). A quantum mirrors is called [SPDC,DFG]-QM if is based on the quantum SPDC and DFG phenomena (3)-(5) in order to transform signal photons, characterized by $(\omega_s, \vec{k}_s, \vec{e}_s, \mu_s)$, into idler photons with $(\omega_p - \omega_s, \vec{k}_p - \vec{k}_s, \vec{e}_s^*, -\mu_s) \equiv (\omega_i, \vec{k}_i, \vec{e}_i, \mu_i)$. Therefore, according to the schematic description from Fig. 3, a [SPDC,DFG]-QM is composed from: a high quality *laser pump* (p), *a transparent crystal* in which all the three photon reactions (3)-(5) can be produced, all satisfying the same energy-momentum conservation laws: $\omega_p = \omega_s + \omega_i$, $k_p = k_s + k_i$.

In these conditions a new geometric optics can be developed on the basis of the concept of quantum mirrors (QM) as shown in refs. [9-11]. Hence, the laws of the plane quantum mirror (see Fig.4) and that of spherical quantum mirror (SQM) (see Fig. 5), observed in the ghost imaging experiments [6-7], are proved as natural consequences of the energy-momentum conservation laws. By the quantum mirroring mechanism (see Fig.3) the objects and their images can be considered as being on the same optical paths. Therefore, the key of ghost imaging mystery can be given by the electromagnetic crossing symmetric photon reactions (3)-(5) or more general by DFG-phenomena (see Fig.1). Indeed, in the case of ghost imaging observed in the papers [6,7] the ghost image can be produced as follows (see Fig.3): The image forms indirectly from the signal photons that come out of the flash, bounce off the aperture, and then are focused through the lens onto QM where they are transformed in idler photons $i_s$ which are collected in the idler detector. So, in this case, the image is not formed directly from signal photons that hits the aperture and bounces back. The image is formed only by idler $i_s$-photons from QM-sources that did not hit the object, but are obtained via crossing symmetric photon reaction (4) [or equivalently, as frequency-difference generation (DFG)] and not via photon reaction (3). Therefore, the crossing symmetry is the heart of ghost imaging, ghost diffraction and ghost interference, phenomena.

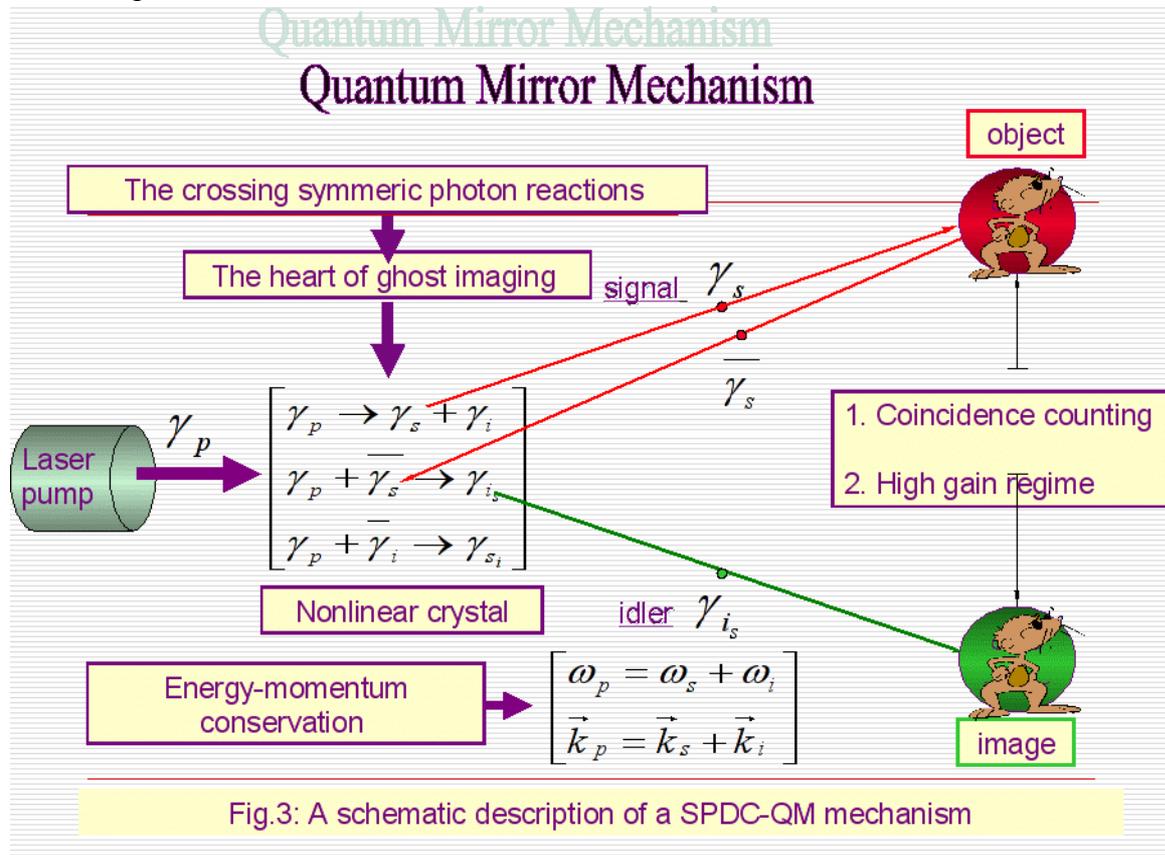

Fig.3: A schematic description of a SPDC-QM mechanism

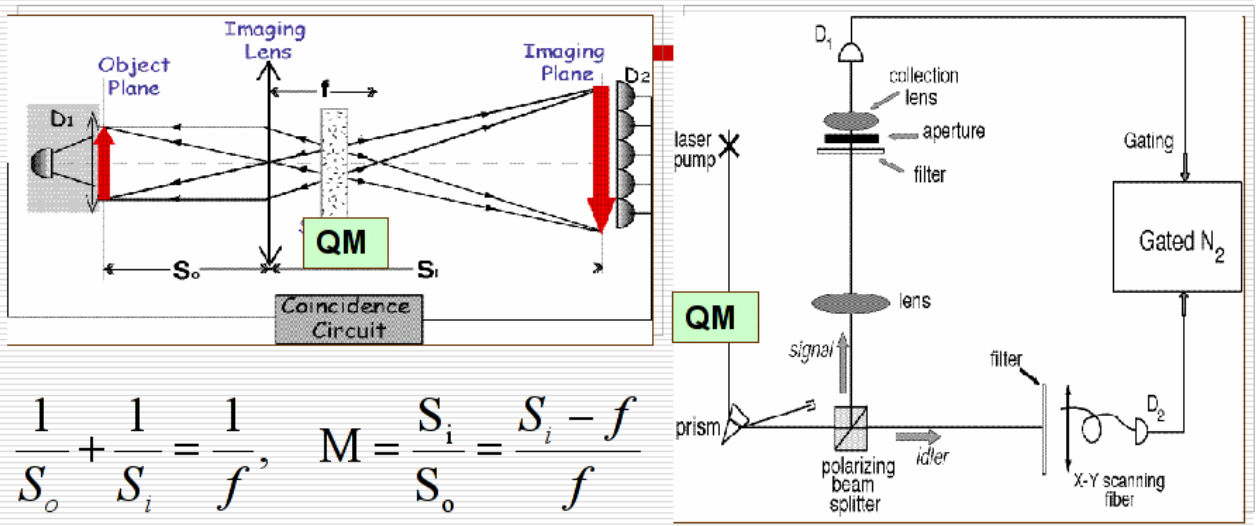

The proof is based on the similar triangles from which we obtain the formulas for M.

Fig.4: Law of the thin lens in the ghost imaging scheme where the PQM-source behaves just as a mirror. The Quantum Mirrors positions are shown by QM.

Clearly, a SPDC crystal illuminated by a high quality laser beam can acts as real quantum mirrors since by the crossing processes (4) (or (5)) a signal photon (or idler photon) is transformed in an idler photon $i_s$ (or signal photon $s_i$ ), respectively.

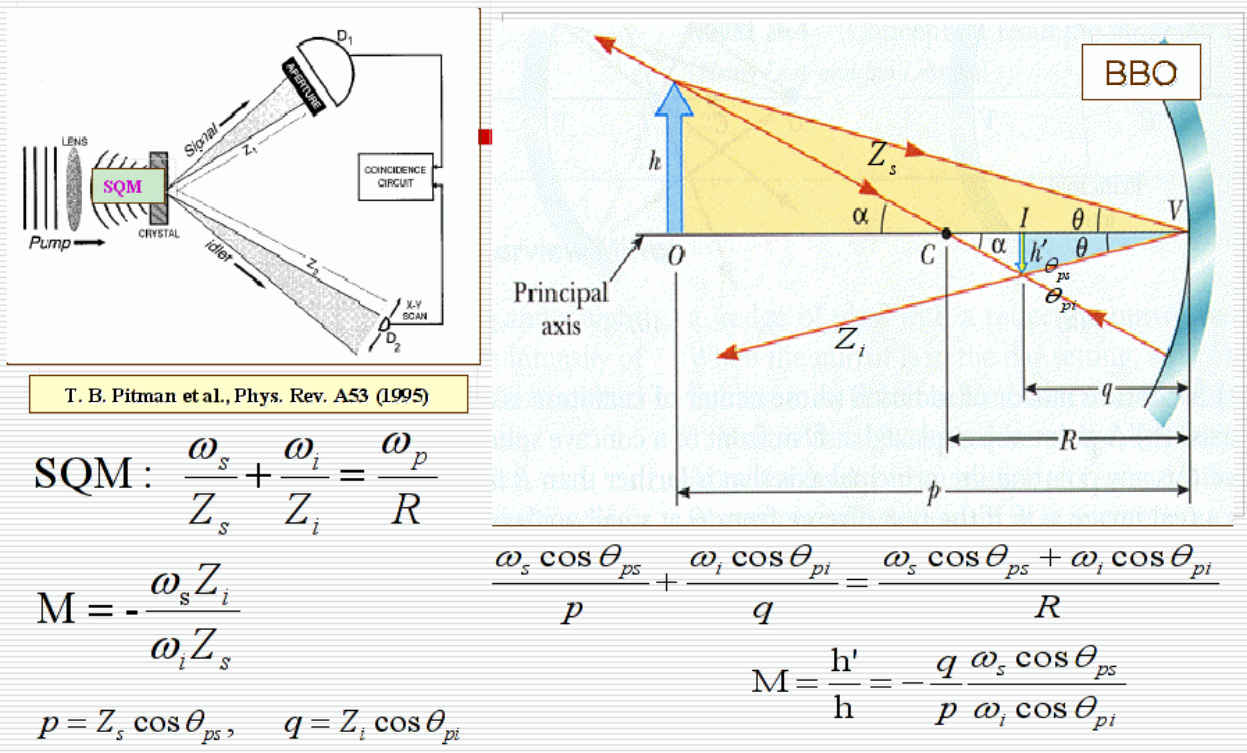

Fig. 5: Experimental verification of the Spherical Quantum Laws

### Entanglement is not necessary for ghost phenomena

Quantum: **Pittman et al.** [Phys. Rev. A 52, R3429-R3432 (1995)]
- performed an imaging experiment with quantum SPDC-light
- with photon-counting bucket and pinhole detectors plus coincidence counting electronics
- and obtained an image without background
- interpreted as a phenomenon due to the *quantum entanglement*

Quantum: **Strekalov et al.** [Phys. Rev. Lett. 74, 3600 (1995)].
- performed an interference experiment with SPDC-light
- with photon-counting bucket and pinhole detectors plus coincidence counting electronics
- the experimental result exhibits a typical double-slit interference fringe
- interpreted as a phenomenon due to the *quantum entanglement*

Classical: **Valencia et al.** [ Phys. Rev. Lett. 94 (2005)] and **Ferri et al.** [ Phys. Rev. Lett. 94 (2005)].
- Performed an imaging experiment with pseudo-thermal light
- with photon counting (Valencia) or a CCD camera (Ferri)
- plus coincidence counting (Valencia) or correlation (Ferri)
- and obtained an image with background
- showing that the *entanglement is not necessary for ghost imaging*

Classical: **Lu Gao et al.** [Optics Commun. 281, 2838-2841 (2008)]
- performed an interference experiment with pseudo-thermal light
- illuminates two spatially separated apertures
- the experimental result exhibits a typical double-slit interference fringe
- showing that the *entanglement is not necessary for the ghost* interference

Fig.6: A short description of the main experimental results obtained in the debate Classical-Quantum nature of ghost phenomena

The quantum mirrors can be "plane"[11] (PQM) and "spherical quantum mirrors" (SQM) (see figs.3-4) according with the character of incoming laser waves (plane waves or spherical waves). In order to avoid many complications, in this presentation we will work only in the thin crystal approximation and only for SPDC-QM. Now, it is important to note that using the QM-concepts [9-12] the results PQM and SQM from Fig.4-5 are obtained only as a consequence of the energy-momentum conservation (or phase matching conditions) without any kind of photon entanglement. In order to illustrate this we present in Figs. 5 a proof of SQM-law using only energy-momentum conservation law.

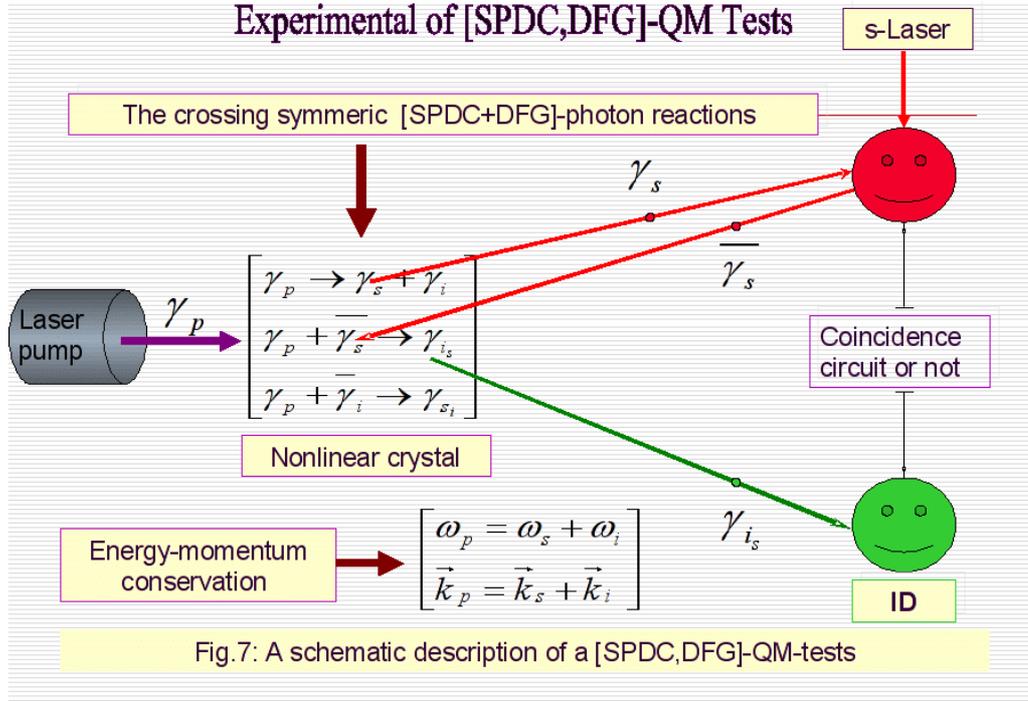

Fig.7: A schematic description of a [SPDC,DFG]-QM-tests

Also, it is important to remark that, the very high selectivity of the SPDC-QM is given by the fact that the energy-momentum conservation laws acts as a daemon which selects only the imaging photons $i_s$ which are produced by the crossing symmetric photon reaction (4) [or more general, in the (DFG), in electromagnetic transitions which are crossing symmetric].

### Experimental Tests of [SPDC-DFG]-Quantum Mirror

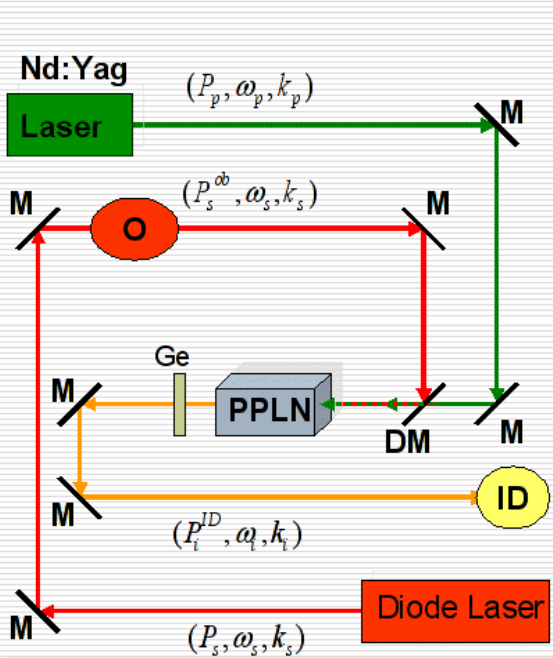

(1) $P_i^{ID} = \dfrac{4\omega_i^2 d_{eff}^2}{\pi\varepsilon_0 n_i n_s n_p} \dfrac{L}{w_{0s}^2 + w_{0p}^2} \dfrac{h(\mu,\xi)}{b} P_p P_s^{ob}$,

(2) $h(\mu,\xi) = \dfrac{1}{2\xi}\displaystyle\int_0^\xi\int_0^\xi \dfrac{e^{-i(\tau-\tau')\Delta k b/2}}{1-\dfrac{i}{2}\left[\dfrac{1+\mu}{1-\mu}-\dfrac{1-\mu}{1+\mu}\right](\tau-\tau')+\tau\tau'} d\tau d\tau'$

where $\mu = k_s/k_p$, focusing parameter $\xi = L/b$ and
the phase-mismatch: $\Delta k = k_p - k_s - k_i$.
Here, $k_{i,p,s}$ are the wave vectors, $P_p$ the pump laser power
$w_{0s,0p}$ the beam waist of pump and signal beams inside
the crystal, L = 50mm the crystal lenght, and $b = b_{i,p,s} = $
$= \dfrac{n_{i,p,s}\omega_{i,p,s}}{c}w_{0i,0p,0s}^2 = 24mm$ the cofocal parameter of
the laser beam involved. In the case of quasi-phase matching
the effective nonlinear coeficient is $d_{eff} = 27(2/\pi)M_{ij} pm/V$,
where $M_{ij} = 0.85$ the Miller factor for DFG process.
[see: S. Stry et al.[13], and Fig.2]

Fig. 8: Schematic of description of the experimental test of the ghost imaging via [SPDC-DFG]-QM (see Fig.2). M-plane mirrors, DM-dichroic mirror, PPLN-periodically poled lithium niobate crystal, Ge-germanium window, O-object, and ID-image detector, $P_s^{ob}$ is the s-power transmited by object, while $P_i^{ID}$ is i-power receeived by ID-detector.

Eq. (1) from Fig.8 shows typical features for a three-wave parametric mixing process:
1. So, the i-power $P_i^{ID}$ received by ID-detector is proportional to the nonlinear optical figure of merit, $d_{eff}^2/(n_i n_s n_p)$.
2. The output idler power varies linearly with the product of the input powers $P_s^{ob} \cdot P_p$
3. The i- power $P_i^{ID}$ is proportional to the square of the idler frequency $\omega_i$.
4. The above idler power varies with the crystal length $L$ in the case of Gaussian beam coupling, and reaches a maximum value with an optimum focusing parameter of $\xi \approx 1.3$. The h-function reduces to $h = \xi$ when using loose focusing parameter $\xi \ll 1$, which makes the DFG power proportional to $L^2$, as in the case of the plane-wave approximation where the nonlinear generation power at the resultant idler frequency $\omega_i$, in the non absorptive medium, can be written as:

$$P_{iID} \propto \frac{d_{eff}^2 L^2}{n_i \cdot n_s \cdot n_p} \cdot \omega_i^2 \cdot P_p \cdot P_{sob} \cdot \left[\frac{\sin(|\Delta k| L/2)}{|\Delta k| L/2}\right]^2 \quad (8)$$

Therefore, the image visibility in all ghost imaging phenomena can be controlled by varying the above essential parameters $P_{sob} \cdot P_p$ and $d_{eff}^2/(n_i n_s n_p)$, from Eq. (8).

On the other hand is well known that Souto Riberiro et al. [23] performed the Young's double-slit experiment with light produced in stimulated down conversion. They studied the *degree of visibility* of the interference patterns as a function of the *mean occupation number per mode N*, or the inducing laser intensity $I_s$, and they demonstrated that the spatial coherence in the signal beam can be controlled by means of its conjugate pair. (See Figs. 9a,b for the experimental results). We prove that QM-mechanism can explain completely all these very important results. The results presented in Fig. 9a,b strongly suggest that similar results can be obtained for ghost image. So, on the base of the results presented in Fig.8 visibility of a ghost image ID can be controlled with the intensity $I_s$ of the signal laser. So, with such an experiment we can finally prove that the coincidences are not necessary for ghost phenomena (imaging, diffraction and interferences) in agreement with the conclusion that the *entanglement is not necessary for ghost imaging, ghost diffraction and ghost interferences phenomena* [see experiments quoted in Fig.6].

## 4. Discussions and Conclusions

The results and conclusions obtained in this paper can be summarized as follows:

(i) Quantum electrodynamics (QED) is one of the working example of the field theory which is indeed crossing symmetric [1,2]. This property of the QED is experimentally verified [3] with very high accuracy.

(ii) The quantum mirror (QM) property of the SPDC-source (laser pump+crystal) is clearly produced by the crossing symmetric photon reactions (4)-(5) [or DFG-crossing symmetric transitions in the source of thermal light, etc.] These QM-instruments possesses some peculiar properties, such as:

  (a) The *angle of incidence* $\theta_{ps}(\omega_s)$ is not equal with the *angle of reflection* $\theta_{pi}(\omega_i)$, excepting the degenerate case $\omega_s = \omega_i$. These angles must obey the transverse momentum conservation law $\omega_s \sin\theta_{ps} = \omega_i \sin\theta_{pi}$ at crystal surface;
  (b) A nondegenerate QM, producing DFG, change the color of incident s-beam in the color of the "reflected" $i_s$-photon beam;
  (c) A QM-optical instrument possess a *high photon-selectivity* since only the **s**-photons which satisfy the *energy-momentum conservation laws* $\omega_p - \omega_s = \omega_i$, $\vec{k}_p - \vec{k}_s = \vec{k}_i$ are "reflected" as $\mathbf{i_s}$ –photons;
  (d) The QM as an independent optical instrument can be combined with other optical classical devices. Hence, the laws of the plane quantum mirror (see Fig.4) and that of spherical quantum mirror (SQM) (see Fig.5), observed experimentally in the ghost imaging experiments [6-7], are obtained as natural consequences of the energy-momentum conservation laws;

(e) Substantial improvements of the optical devices (e.g. telescopes, microscopes, etc.), are expected to be obtained by using quantum mirrors (see [9]-[12]) instead of the usual mirrors. Some geometrical advantages (e.g., the magnifications and distances control by varying the pump wavelength, etc.) as well as signal processing advantages (e.g., the high resolution, the high fidelity and amplification of the incoming beam intensity, distortion undoing for the signal rays, coherence preserving, etc.) are expected to be obtained by using the quantum mirror instead of the usual mirrors.

(iii) In the ghost-imaging experiments (see Figs.3-5) the image forms indirectly from the signal photons that come from the object, and are focused through the lens onto QM where they are transformed in idler photons $i_s$ which are collected in the image detector [or on a photo-reactive film, etc.]. So, in this case, the image is formed only by idler $i_s$-photons from QM-sources that did not hit the object (see again Fig.3), but are obtained via crossing symmetric photon reaction (4).

(iv) The recent results [19]-[22] definitely proved experimentally that the *entanglement is not necessary in ghost imaging*. These important results are summarized in Fig. 6. Consequently, we suspected that measurements in coincidence counting regime, can be used only for the background subtractions, just as in nuclear and elementary particle physics. An experimental test of the <u>SPDC-QM</u> is suggested in Fig.7.

(v) An fundamental more efficient experimental test of the <u>DFG-QM</u> is proposed in Fig.8. The DFG typical features expressed by Eq. (1), in Fig. 8, allowed us to conclude that *image visibility* of all ghost phenomena can be controlled by varying the following parameters: $P_{s_{ob}} \cdot P_p$ and $d_{eff}^2/(n_i n_s n_p)$, from Eq.(1) in Fig.8. This conclusion is in excellent agreement with the experimental results of Souto Ribeiro et al. [23] presented in Figs.9a,b. We note of course that a high quality of a ghost image will be obtained by illuminating the object with a high intensity s-laser. Finally, we believe that the results obtained here are encouraging for new theoretical and experimental investigations since the crossing symmetry as heart of ghost (imaging, interference-diffraction) phenomena can be of great importance not only from fundamental point of view but also for practical applications.

## Control of Young's fringes visibility by stimulated down-conversion

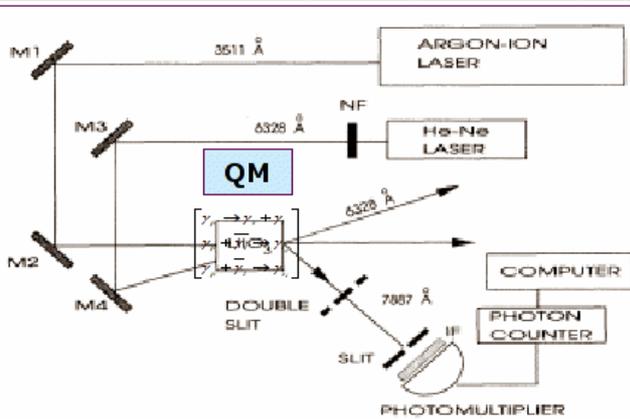

FIG.1. Schematic diagram of the experimental setup for Young's double-slit experiment. $M_1, M_2, M_3, M_4$ are mirrors, IF is an interference filter, and NF is an neutral filter.

Performing the double-slit Young experiment, Souto Ribeiro et al. [23] shown that the interference visibilities can be well controlled by varying the induced s-laser intensity. The results indicate that down conversion light statistics change during this variation. So the best visibilities of Young's fringes are obtained with very high intensities of the induced laser intensities. The results can be completely explained via QM-mechanism [see Eq. (1), Fig.7]

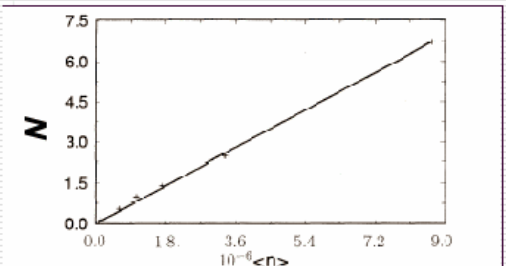

FIG. 3. Experimental occupation numbers $N$ as a function of $\langle n \rangle$ the inducing laser mean photon number. A fit to Eq. (8) gives $\beta = (7.74 \pm 0.11) \times 10^{-7}$. Errors bars are the same size as the symbols.

$$N = \beta \langle n \rangle \quad (8)$$

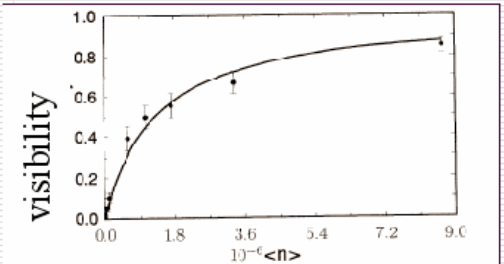

FIG. 4. Experimental results (circle) and theory (line) for the Young's fringes visibilities as a function of the mean photon number $\langle n \rangle$ of the inducing laser.

Fig.9a: Control of Young's visibility by stimulated down conversion. Position of the quantum mirror is given by QM.

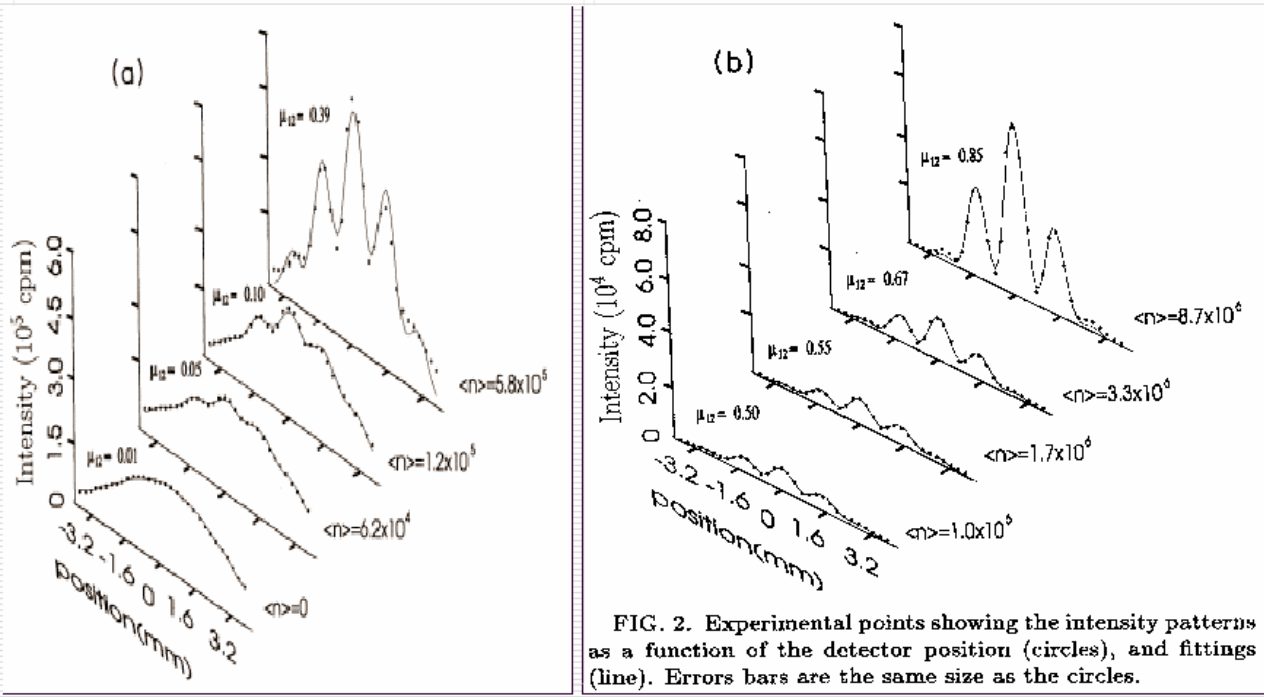

FIG. 2. Experimental points showing the intensity patterns as a function of the detector position (circles), and fittings (line). Errors bars are the same size as the circles.

Fig.9b: Control of Young fringes visibility by stimulated down conversion (see Ref. [23]).

**Acknowledgments**

This research was supported by CNCSIS under contract ID-52-283/2007 and also by CNMP project 71-nr.131/2007.